\documentclass[journal]{IEEEtran}
\usepackage{enumitem}
\usepackage{xcolor}
\usepackage{amsmath,amsfonts}
\usepackage{algorithm}
\usepackage[group-separator={,},group-minimum-digits=4]{siunitx}
\usepackage{threeparttable}
\usepackage{array}
\usepackage[caption=false,font=normalsize,labelfont=sf,textfont=sf]{subfig}
\usepackage{textcomp}
\usepackage{stfloats}
\usepackage{url}
\usepackage{verbatim}
\usepackage{graphicx}
\usepackage{algpseudocode}
\usepackage{cite}
\usepackage{balance}
\usepackage{mathrsfs}
\usepackage{bbding}
\usepackage[colorlinks=true,
            linkcolor=blue,
            anchorcolor=blue,
            citecolor=blue]{hyperref}

\begin{document}
\title{\fontsize{23pt}{\baselineskip}\selectfont{
    DISCO Might Not Be Funky: Random Intelligent Reflective Surface Configurations That Attack
} 
}
\author{
{       Huan~Huang,~\textit{Member, IEEE,}
        Lipeng~Dai,
        Hongliang~Zhang,~\textit{Member, IEEE,}
        Chongfu~Zhang, 
        Zhongxing~Tian,
        Yi~Cai,~\textit{Senior~Member,~IEEE,}
        A.~Lee~Swindlehurst,~\textit{Fellow,~IEEE,}
        and~Zhu~Han,~\textit{Fellow,~IEEE}
         
}% <-this % stops a space
\vspace{-0.45cm}
\thanks{
    This work was supported by the National Natural Science Foundation of China (62250710164, 62275185, 62371011), and partially supported by the U.S. National
    Science Foundation (CNS--2107216, CNS--2128368, CNS--2107182, CMMI--2222810, ECCS-2302469, ECCS--2030029), 
    US Department of Transportation, Toyota and Amazon. (\textit{Corresponding author: Huan~Huang})
 
H.~Huang, Z.~Tian, and Y.~Cai are with the School of Electronic and Information Engineering, 
Soochow University, Suzhou 215006, China 
(e-mail: hhuang1799@gmail.com, zxtian@ieee.org, yicai@ieee.org).

L.~Dai and C.~Zhang are with the School of Information and Communication Engineering, 
University of Electronic Science and Technology of China, Chengdu 611731, China 
(e-mail: dlp1022@163.com, cfzhang@uestc.edu.cn).

H.~Zhang is with the State Key Laboratory of Advanced Optical Communication Systems and Networks, 
School of Electronics, Peking University, Beijing 100871, China 
(e-mail: hongliang.zhang92@gmail.com).

A.~L.~Swindlehurst is with the Center for Pervasive Communications and Computing, 
University of California, Irvine, CA 92697, USA (e-mail: swindle@uci.edu).

Z.~Han is with the University of Houston, Houston, TX 77004 USA 
(email: hanzhu22@gmail.com).

}
}

\maketitle
\begin{abstract}
Emerging intelligent reflective surfaces (IRSs) significantly improve system performance, 
but also pose a significant risk for physical layer security (PLS).
Unlike the extensive research on legitimate IRS-enhanced communications, 
in this article we present an adversarial IRS-based fully-passive jammer (FPJ).
We describe typical application scenarios for Disco IRS (DIRS)-based FPJ,
where an illegitimate IRS with random, time-varying reflection properties acts like a ``disco ball" 
to randomly change the propagation environment.
We introduce the principles of DIRS-based FPJ and overview existing investigations of the technology, 
including a design example employing one-bit phase shifters.
The DIRS-based FPJ can be implemented without either jamming power or 
channel state information (CSI) for the legitimate users (LUs).
It does not suffer from the energy constraints of traditional active jammers, nor does it
require any knowledge of the LU channels. 
In addition to the proposed jamming attack, we also propose an anti-jamming strategy 
that requires only statistical rather than instantaneous CSI.
Furthermore, we present a data frame structure that enables the legitimate access point (AP) to estimate 
the DIRS-jammed channels' statistical characteristics in the presence of the DIRS jamming.
Typical cases are discussed to show the impact of the DIRS-based FPJ 
and the feasibility of the anti-jamming precoder (AJP). 
Moreover, we outline future research directions and challenges for the DIRS-based FPJ and its anti-jamming precoding
to stimulate this line of research and pave the way for practical applications.

\end{abstract}


\section{Introduction}\label{Intro}
Due to the broadcast and superposition nature of wireless channels, 
the open wireless air interface is vulnerable to malicious attacks 
such as jamming or denial-of-service attacks~\cite{PLSsur1,DoSsur1}. 
Jamming attacks can be launched to intentionally disrupt wireless communication networks 
such as Wi-Fi, Bluetooth, Internet of Things (IoT),
and cellular networks.
In traditional wireless systems, active jammers (AJs), which impose 
intentional interference on the communication between 
an access point (AP) and its legitimate 
users (LUs), have been widely investigated. 
In general, physical-layer AJs can 
be classified as constant AJs, intermittent AJs, reactive AJs, 
and adaptive AJs~\cite{PLSsur1}.
A constant AJ continuously broadcasts jamming signals, 
such as pseudorandom noise or Gaussian-modulated waveforms, 
over the wireless air interface to prevent the AP from communicating with the LUs. 
However, constant AJs are energy-inefficient because they constantly consume power. 
To address this issue, intermittent AJs, reactive AJs, and adaptive AJs have been proposed whose
basic motivation is to reduce the duration of the jamming transmission 
and hence reduce the energy consumption. However, all types of active jamming require a certain amount of energy consumption to effectively attack the LUs.

Recently, intelligent reflective surfaces (IRSs)~\cite{IRSsur1,IRSsur2,IRSsuradd}, 
which reflect electromagnetic waves in a controlled manner, 
have been proposed as a promising technology for future 6G systems. 
An IRS is an ultra-thin surface equipped with multiple subwavelength reflective elements 
whose electromagnetic responses (e.g., amplitudes and phase shifts) can be configured, for instance, 
by simple programmable PIN or varactor diodes. 
Previous works have mainly focused on using legitimate IRSs to improve performance, 
assuming that the legitimate AP knows the IRS-related channel state information (CSI), 
and can control their phase responses.
At the same time, a handful of works have examined the risk 
that illegitimate IRSs pose to physical layer security (PLS)~\cite{IIRSSur1,IIRSSur2}. 
For example, the authors in~\cite{IIRSSur1} illustrated that illegitimate IRSs can be used to help 
the AJs enhance their jamming attacks, especially in the case of AJ-LU link blocking. 
However, illegitimate IRS-aided AJs also have the inherent disadvantage of requiring significant energy consumption. 
Considering this inherent energy consumption drawback, \emph{can jamming attacks be launched without jamming power?}

The authors in~\cite{PassJamSU} have reported an adversarial IRS-based passive jammer (PJ) 
for single-user multiple-input single-output (SU-MISO) systems that essentially consumes 
no power. This adversarial IRS destructively adds the reflected path signal to  
the direct path signal to minimize the received power at the LU. 
Although this PJ can launch jamming attacks without consuming power, 
the CSI of all wireless channels involved must be known at the unauthorized IRS. 
Due to the passive nature of IRSs, the CSI of IRS-related channels is estimated jointly with that of the AP and LUs. 
In other words, if the illegitimate IRS aims to acquire LU CSI, 
it must train to learn CSI jointly with the legitimate AP and LUs. 
Therefore, the assumption that the illegitimate IRS knows the CSI of all channels is unrealistic. 
Given the difficulty of illegitimate IRSs in acquiring CSI, \emph{can jamming attacks be launched without either jamming power or LU CSI?}

The emergence of IRSs in support of existing wireless systems significantly improves their performance without noticeably increasing the power consumption or cost. 
However, it also poses a significant risk for PLS. 
For example, the authors in~\cite{MyGC23Extension} have summarised some typical IRS-based attack strategies such as IRS-based jamming and eavesdropping.
In this article,  we explore the potential of illegitimate IRSs
in launching fully-passive jamming attacks and present a new attack approach referred to as Disco IRS (DIRS)-based fully passive jammer (FPJ).
We discuss the principle, advantages, and implementation of this new attack. 
In view of the great threat that such fully-passive attacks pose to communication networks, 
we further discuss an anti-jamming precoding strategy to counteract it. 
Moreover, we outline future research directions and challenges.

\section{Disco IRS Based Active Channel Aging: A New Jamming Attack}\label{FPJ}
\textbf{Fundamentals:} To implement an FPJ without relying on either jamming power or LU CSI, 
the interesting idea of DIRS was first proposed in~\cite{DIRSVT}, 
which described an illegitimate IRS with random reflection properties that acts like a ``disco ball."  
 It is worth noting that the DIRS-based FPJ does not require jamming power, 
but it does consume a relatively small amount of circuit power to control its operation. 
While a disco ball makes dancing more funky, in a wireless communication system the result is not so pleasing.
In a DIRS system, the controller generates a single realization of random 
independently and identically distributed (i.i.d.) phase shifts once during 
the \emph{pilot transmission (PT)} phase, and then a different i.i.d. set of phase shifts
during the subsequent \emph{data transmission (DT)} phase.
As a result, serious active channel aging (ACA) interference, i.e., a type of inter-user interference (IUI), is introduced.
Note that the DIRS-based ACA is different from the channel aging (CA) in traditional MU-MISO systems, 
which is caused by channels that vary between when they are learned at the legitimate AP and when they are used for precoding due to time variations in the channel and delays in the computation~\cite{ChanAge}.

\begin{table*}
    \footnotesize
    \centering
    \caption{Comparison of Different Jammers}
    \label{tab1}
    \begin{threeparttable}
    \begin{tabular}{ |c|c|c|c|c| }
    \hline
    Category                                             &Jamming energy       &Channel knowledge           &MIMO-based cancellation     &Frequency-hopping/spread spectrum  \\
    \hline
    Active jammer (AJ)~\cite{PLSsur1}                    &Required                 &Not Required                    &\textcolor[rgb]{0.00,1.00,0.00}{$\surd $}                       &\textcolor[rgb]{0.00,1.00,0.00}{$\surd $} \\
    \hline
    IRS-aided AJ~\cite{IIRSSur1}                         &Required                 &Required                        &\textcolor[rgb]{0.00,1.00,0.00}{$\surd $}                       &\textcolor[rgb]{0.00,1.00,0.00}{$\surd $}  \\                                 
    \hline
    Passive jammer (PJ)~\cite{PassJamSU}                 &Not Required             &Required                        &\textcolor[rgb]{0.00,1.00,0.00}{$\surd $}                       &\textcolor[rgb]{1.00,0.00,0.00}{$\times $}  \\                                 
    \hline
    Fully-passive jammer (FPJ)~\cite{DIRSVT,DIRSTWC}     &Not Required             &Not Required                    &\textcolor[rgb]{1.00,0.00,0.00}{$\times $}                      &\textcolor[rgb]{1.00,0.00,0.00}{$\times $}  \\                                 
    \hline
    \multicolumn{4}{l}{\scriptsize The mark {\textcolor[rgb]{0.00,1.00,0.00}{$\surd $}} represents 
    that the scheme works;} \\
    \multicolumn{4}{l}{\scriptsize The mark \textcolor[rgb]{1.00,0.00,0.00}{$\times $} represents that the scheme does not work.} 
    \end{tabular}
    \end{threeparttable}
\end{table*}

Based on this ACA idea~\cite{DIRSVT}, the work in~\cite{DIRSTWC} further illustrated that the DIRS-based ACA interference 
can also be introduced by turning off the illegitimate IRS (i.e., the wireless signals are perfectly absorbed by the DIRS) during the \emph{PT} phase
and then generating i.i.d. random reflecting vectors multiple times during the \emph{DT} phase.
It is worth noting that such a temporal DIRS-based FPJ must know when the \emph{PT} phase ends 
and the \emph{DT} phase begins, which requires some synchronization with the legitimate system.
The DIRS-based FPJs in~\cite{DIRSVT,DIRSTWC} which require neither jamming power nor LU CSI, 
pose a significant risk to PLS.
For example, the theoretical analysis in~\cite{DIRSTWC} showed that a DIRS-based FPJ 
using only one-bit quantized phase shifts can achieve the desired jamming effects 
as long as the number of DIRS reflective elements is large enough. 
The immediate question, therefore, is \emph{how to mitigate the DIRS-based ACA interference}~\cite{DIRSTWC}.

To address this issue,~\cite{MyGC23} first proposed an anti-jamming precoder (AJP) to counteract
 the temporal DIRS-based FPJ.
In particular, the statistical characteristics of the DIRS-jammed channels 
were derived and an AJP that can achieve 
the maximum signal-to-jamming-plus-noise ratio (SJNR) was developed. 
The work in~\cite{MyGC23Extension} showed how the legitimate AP can acquire the statistical characteristics 
in a practical way, 
and also extended the AJP in~\cite{MyGC23} to address persistent 
DIRS-based fully-passive jamming. Since the DIRS-based FPJs in~\cite{DIRSVT} and~\cite{DIRSTWC} are 
included in the persistent DIRS-based FPJ model, the  AJP
is suitable for all DIRS-based fully-passive attacks.

\textbf{DIRS Features:} 
The above prior work also derived some interesting properties of this temporal DIRS-based ACA 
via theoretical analysis. 
For example, the jamming impact of the DIRS-based FPJ proposed in~\cite{DIRSTWC} cannot be mitigated 
by increasing the transmit power, 
and classical anti-jamming approaches, such as spread spectrum and frequency-hopping techniques, 
cannot be used against an FPJ because the source of the jamming attacks is
the transmit signals themselves and has the same characteristics (e.g., carrier frequencies, etc). 

Multi-input multi-output (MIMO) interference cancellation 
has also been studied as an important anti-jamming approach~\cite{JammResilientCommun}. 
However, MIMO interference cancellation is effective for DIRS-based ACA interference 
only if the legitimate AP has knowledge of the LU and DIRS-jammed channels.
Since the DIRS is passive and the phase shifts and amplitudes are randomly generated, 
the DIRS-based ACA interference cannot be mitigated by MIMO interference cancellation.
Table~\ref{tab1} compares the characteristics of AJs, PJs, and FPJs.

\begin{figure*}[!t]
    \centering
    \includegraphics[scale=0.5]{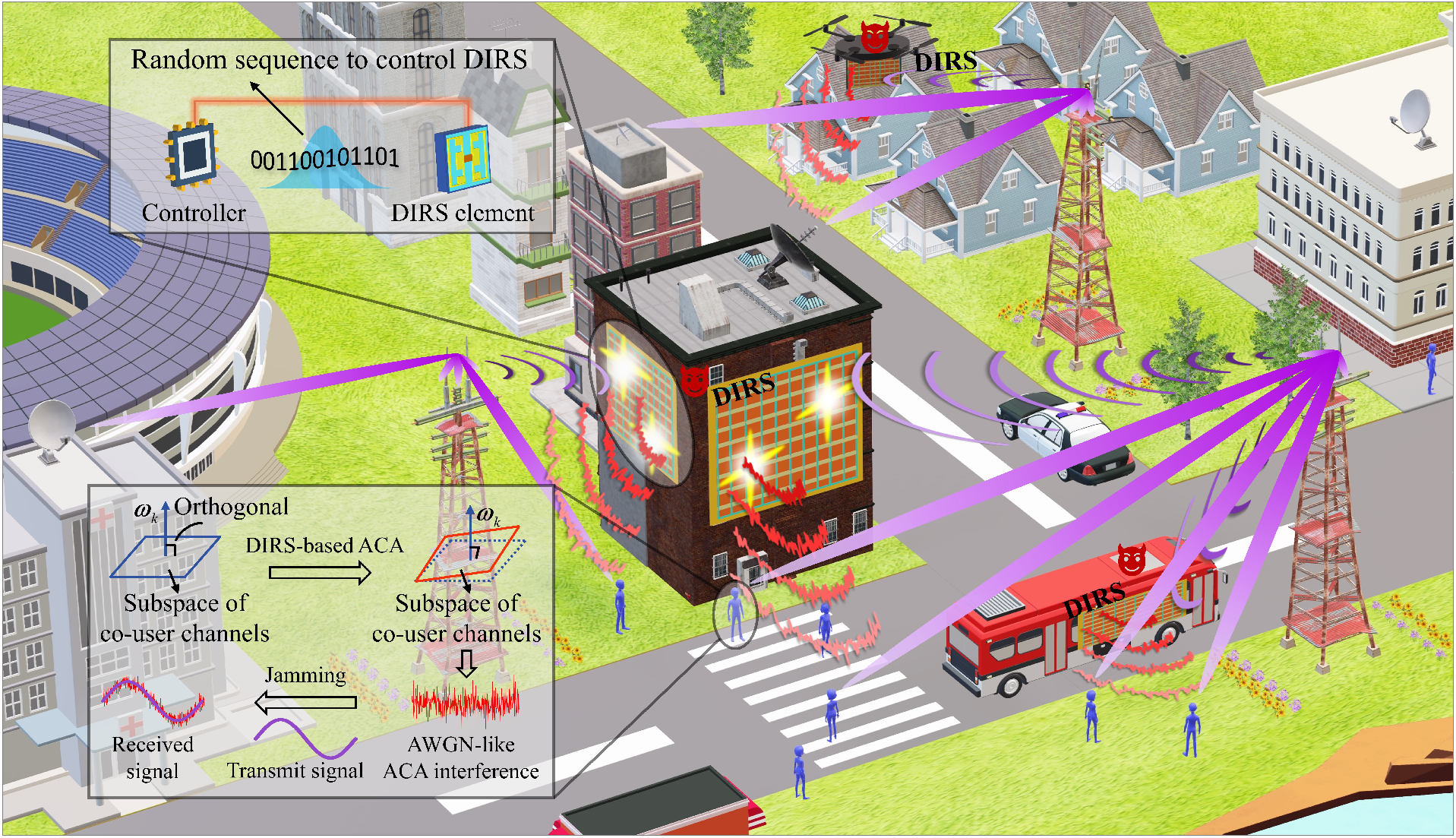}
    \caption{Implementation of disco intelligent reflective surface (DIRS) based fully-passive jamming attacks, where the DIRS reflection properties, i.e., the phase shifts and amplitudes are randomly and independently generated by the DIRS controller.}
    \label{fig1}
\end{figure*}

{\textbf{DIRS Applications:}} Fig.~\ref{fig1} illustrates different types of wireless communications systems 
that can be jammed by persistent DIRS-based FPJs. 
Note that there are many possible deployment strategies for DIRSs, for instance near a legitimate AP,
on an unmanned aerial vehicle (UAV)~\cite{IRSsuradd}, or on public transportation vehicles.
The existing works~\cite{DIRSVT,DIRSTWC,MyGC23,MyGC23Extension} 
have only investigated fixed DIRS deployments
near the legitimate AP, and DIRS placement on mobile platforms (e.g., UAVs) 
is worthy of further investigation, including optimization of the UAV-based DIRS route, etc.

{\textbf{DIRS Principles:}} In a persistent DIRS-based FPJ, a random sequence 
with i.i.d. random elements generated by 
a  controller is used to adjust the DIRS phase shifts.
While the IRS phase shifts is controlled, 
the amplitudes of the reflective elements are a function of their corresponding phase shifts 
due to the unique electromagnetic properties of typical IRS elements~\cite{IRSsur1}.

In an MU-MIMO/MISO system, the legitimate AP jointly trains the CSI with the LUs during the \emph{PT} phase, 
and the CSI is used to design a precoder for transmitting signals to the LUs during the \emph{DT} phase. 
The LU CSI can be jointly estimated using existing algorithms, such as for example the least squares (LS) algorithm.
In general, wireless channels are assumed to remain unchanged during a channel coherence interval 
consisting of a \emph{PT} phase followed by a  \emph{DT} phase. 
Based on this assumption, the legitimate AP can design a transmit precoder based on the CSI obtained from the \emph{PT} phase.
However, the programmable IRS provides the ability to actively age the wireless channels within their coherence interval 
and thus produces a situation where the CSI obtained in the \emph{PT} phase is different from that in the \emph{DT} phase.

Inspired by this idea, a DIRS is introduced to actively age the wireless channels~\cite{DIRSVT,DIRSTWC}. 
Specifically, assume that the length of the \emph{PT} phase is $T_{\!P}$ and that of the \emph{DT} phase is $T_{\!D} = C T_{\!P}$, i.e., 
the length of a channel coherence interval is $T_{\!C} = T_{\!P} + T_{\!D} = \left(C+1\right)T_{\!P}$. 
We can exploit the DIRS to rapidly age the wireless channels, 
and effectively produce a channel with coherence interval much less than $T_{\!C}$. 
As a result, serious DIRS-based ACA interference is introduced, and the LUs are then jammed.
In a persistent DIRS-based FPJ, the DIRS controller generates a single i.i.d. random sequence 
used to tune the reflecting phase shifts during the \emph{PT} phase, 
and then generates different i.i.d. random sequences 
to change the reflecting phases shifts $Q$ ($Q\ge C$) times during the \emph{DT} phase.
If the DIRS controller sets the DIRS reflecting vector to zero during the \emph{PT} phase, 
 this corresponds to the case investigated in~\cite{DIRSTWC}. Moreover, if we let $Q = 1$ during the \emph{DT} phase, 
 the persistent DIRS-based FPJ reduces to the one studied in~\cite{DIRSVT}. 

Taking a legitimate AP with the widely-used ZF transmit precoder as an example, 
in a traditional MU-MISO/MIMO system,
the AP calculates the ZF precoder based on the CSI obtained in the \emph{PT} phase to transmit signals during the \emph{DT} phase. 
If the wireless channels remain unchanged in a channel coherence interval, 
the transmit precoding vector ${\boldsymbol{w}}_{k}$ for the $k$-th LU 
is always orthogonal to the subspace of the other co-channel users, as shown in Fig.~\ref{fig1}. 
However, when the DIRS-based ACA interference is introduced, the wireless channels during 
the \emph{DT} phase are not the same as those during the \emph{PT} phase, 
and the transmit precoding vector ${\boldsymbol{w}}_{k}$ is no longer orthogonal to 
the co-channel user subspaces during the \emph{DT} phase.
 As a result, serious ACA interference is introduced by the persistent DIRS-based FPJ.
As illustrated in Fig.~\ref{fig1}, 
DIRSs cause the signals from the DIRS-jammed channels
to behave like additive Gaussian white noise (AWGN)~\cite{DIRSTWC,MyGC23,MyGC23Extension}. 
As a result, the LUs are jammed.
\section{Implementation of Persistent DIRS-Based FPJ Using One-Bit Phase Shifts}\label{OnebitFPJ}
\begin{figure}[!t]
\centering
\includegraphics[scale=0.439]{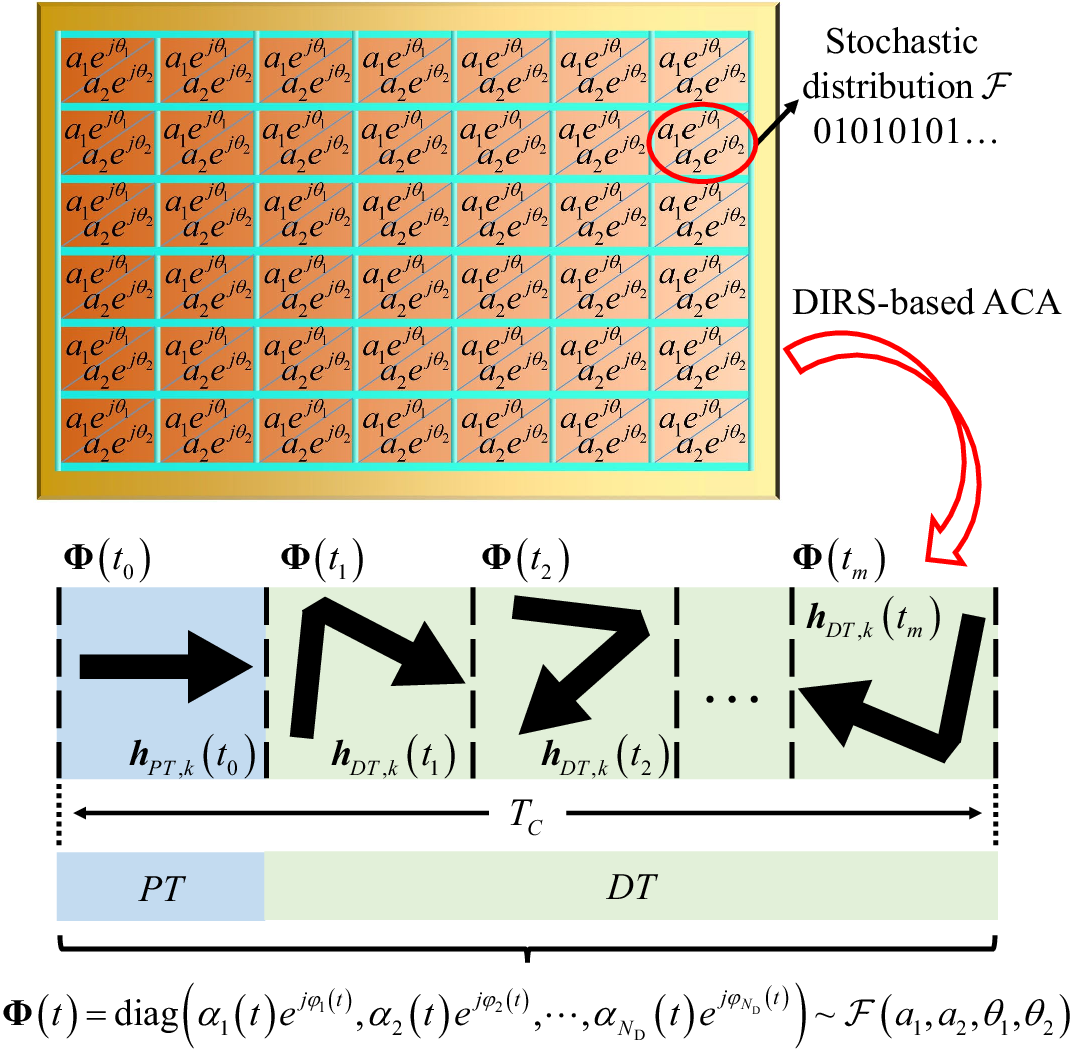}
\caption{Implementation example of a persistent DIRS-based FPJ using an IRS 
with one-bit quantized phase shifts, where the DIRS with random reflection properties actively ages wireless channels within a channel coherence interval.}
\label{fig2}
\end{figure}

An implementation example of a persistent DIRS-based FPJ
using an IRS with one-bit phase shifters is shown in Fig.~\ref{fig2}.
Each reflective element has one-bit quantized phase shifts and
corresponding reflection amplitudes denoted as
$\left\{ {{\theta _1},{\theta _2}} \right\}$ and $\left\{ {{a _1},{a _2}} \right\}$, respectively. 
Furthermore, we assume that the DIRS phase shifts follow the stochastic distribution $\cal F$.
To implement this DISCO approach, the DIRS  controller first generates an i.i.d. random sequence 
following $\cal F$ to control the DIRS phase shifts and amplitudes during the \emph{PT} phase, 
where the diagonal reflecting matrix is denoted by ${\bf{\Phi}}(t_0)$. 
Then, the wireless channel of the $k$-th LU can be written as ${\boldsymbol{h}}_{PT,k}(t_0)$, 
whose CSI is estimated jointly by the legitimate AP and the $k$-th LU during the \emph{PT} phase. 
The DIRS  controller subsequently generates a set of $m$ different i.i.d. random sequences, 
also following $\cal F$,
in order to adjust the DIRS phase shifts and amplitudes during the \emph{DT} phases, 
where the diagonal reflecting matrices are denoted by 
${\bf{\Phi }}({t_1}),{\bf{\Phi }}({t_2}), \cdots ,{\bf{\Phi }}({t_m})$. 
As a result, the $k$-th LU channel is no longer ${\boldsymbol{h}}_{PT,k}(t_0)$ during the \emph{DT} phase, 
but varies randomly according to the random IRS reflections.

Due to the use of the persistent DIRS-based FPJ, the acquired CSI from the \emph{PT} phase is rapidly aged within a channel coherence interval. 
We assume that the DIRS  controller generates i.i.d. random sequences $Q$ times during the \emph{DT} phase. 
In fact, using $C$ different i.i.d. random sequences (i.e., $Q=C$) 
to change the wireless channels $C$ times is enough to shorten 
the original channel coherence from $T_{\!C}$ to $T_{\! P}$, 
in which case there is essentially no time available for data transmission.
The work in~\cite{DIRSTWC} has shown that the DIRS-based ACA interference generated by a one-bit DIRS 
can jam the LU rates to zero as long as the number of DIRS elements is large enough.
As discussed in~\cite{IIRSSur2}, countermeasures based on channel separation 
can only be used to resist AWGN-like ACA interference 
with high multipath resolution such as wideband OFDM. 
However, it is challenging to mitigate the DIRS-based ACA interference for cases 
with low multipath resolution~\cite{IIRSSur2}, such as narrowband systems.

\section{An Anti-Jamming Precoding Strategy for Persistent DIRS-Based FPJ}\label{AntiFPJ}
To design a practical AJP for persistent DIRS-based FPJs, 
legitimate systems must consider the following constraints: 
1) the anti-jamming precoding must be computed without any useful information 
from the illegitimate DIRS; 2) the implementation of the anti-jamming precoding
 cannot require any changes to the existing system architecture. %In other words, the legitimate AP is unable to acquire any useful knowledge to calculate anti-jamming precoder
Fortunately, the pioneering works in~\cite{DIRSTWC,MyGC23,MyGC23Extension} have proved that 
the elements of DIRS-jammed channels converge to a complex Gaussian distribution 
as the number of the DIRS reflective elements is large enough. 
In practice, to cope with the multiplicative large-scale channel fading in cascaded DIRS-jammed channels, 
the DIRS must be equipped with a large number of reflective elements to ensure a significant jamming impact~\cite{DIRSTWC,MyGC23,MyGC23Extension}.
Based on the properties of Gaussian distributions, 
one possible AJP for the $k$-th LU against persistent DIRS-based FPJ is given by
\begin{equation}
{\boldsymbol{w}\!_{{\rm{Anti}},k}} \! \propto \! \max\!.{\rm{eigenvector}}\!\!\left(\!\!\! \frac{{{\boldsymbol{h}_{{\!P\!T},k}}\boldsymbol{h}_{{{\!P\!T}},k}^H \!\!+\! { \delta _k^2 }{{\bf{I}}\!_{N\!_{\rm A}}}}}{{ {{{\widetilde {\bf{H}}}_{{\!P\!T},k}}\widetilde {\bf{H}}_{{\!P\!T},k}^H \!\!+ \!\!\! \left(\!\! {\frac{{{\delta^2}}K}{{{P_0}}} \!+ \!\! {\sum\limits_{u \ne k} \!\! \delta _u^2 }} \!\!\right)\!\!{{\bf{I}}\!_{N\!_{\rm A}}}}  }} \!\!\! \right)  \!,
\label{AntiJamm}
\end{equation}
where $\boldsymbol{h}_{{{\!P\!T}},k}^H$ is the $k$-th LU channel estimated 
during the \emph{PT} phase,
${\delta _k}, k = 1,2,...K$ represents a certain statistical characteristic of 
the DIRS-jammed channel between the AP and the $k$-th LU,
and ${{\widetilde {\bf{H}}}_{{\!P\!T},k}} = \left[{\boldsymbol{h}_{{\!P\!T},1}},\cdots,{\boldsymbol{h}_{{\!P\!T},k-1}},{\boldsymbol{h}_{{\!P\!T},k+1}},\cdots,{\boldsymbol{h}_{{\!P\!T},K}} \right]$
denotes the co-user channels of the $k$-th LU during the \emph{PT} phase. In addition, $P_0$ and $\delta^2$ represent the total transmit power and the variance of the received signals during the \emph{DT} phase.

\begin{figure}[!t]
    \centering
    \includegraphics[scale=0.78]{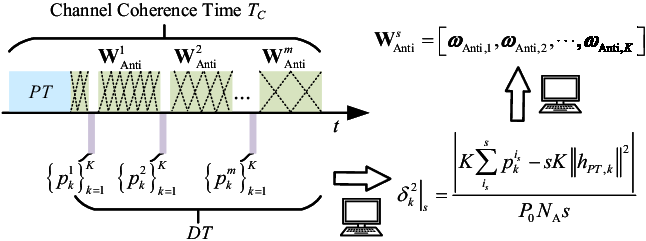}
    \caption{A data frame structure for the legitimate AP to estimate the statistical characteristics of DIRS-jammed channels for the anti-jamming precoder.}
    \label{fig3}
\end{figure}

Note that the implementation of the anti-jamming precoding in~\eqref{AntiJamm} requires 
the statistical characteristics of the DIRS-jammed channels, i.e., $\left\{ {{\delta _k}} \right\}_{k = 1}^K$. 
A feasible data frame structure that can be used by 
the legitimate AP to estimate these statistical characteristics is illustrated in Fig.~\ref{fig3}. 

In the designed frame structure, the LUs only need to feed their received power values back to the AP 
when they detect that they are being jammed, e.g., when they detect a degradation of their SJNRs. 
Only a few bits are required to feed back the received power values 
since they are only scalars.
During a channel coherence interval, we assume that the power information is fed back $m$ times, 
and the $s$-th feedback set of received power values is denoted as
 $\left\{ {p_k^s} \right\}_{k = 1}^K$ ($1 \le s \le m$). 
 Consequently, the $s$-th estimate of the statistical characteristics 
 $\left\{ {{{\left. {{\delta _k^2}} \right|}_s}} \right\}_{k = 1}^K$ can be computed as shown in Fig.~\ref{fig3}.
Then, we can substitute the $s$-th estimate into~\eqref{AntiJamm} to compute the AJP ${\bf{W}}_{{\rm {Anti}},k}^s$.
The work in~\cite{MyGC23Extension} has shown only one or two feedback messages are sufficient to effectively estimate the statistical characteristics for the AJP.

It is seen that the AJP for persistent DIRS-based FPJs has the following interesting properties:
\begin{enumerate}
\item Regardless of the DIRS phase distribution used by the persistent DIRS-based FPJ, 
the proposed AJP is valid 
as long as the number of the DIRS reflective elements is large enough;

\item The legitimate system can acquire the statistical characteristics of 
the DIRS-jammed channels without changing its architecture or 
cooperating with the illegitimate DIRS. 
\end{enumerate}

\section{Case Study}\label{ResDess}
Consider a persistent DIRS-based FPJ case in which an MU-MISO system is jammed by a 2048-element 
one-bit DIRS with phase shifts and amplitudes randomly chosen from $\{\frac{\pi}{9},\frac{7\pi}{6}\}$ 
and $\{0.8,1\}$~\cite{IRSsur1}. The legitimate AP equipped with 16 antennas 
is located at (0 m, 0 m, 5 m) and communicates with 12 single-antenna LUs which 
are randomly distributed in a circular region with a radius of 20 m and centered at (0 m, 180 m, 0 m). 
The DIRS is deployed at (-$d_{\rm{AD}}$ m, 0 m, 5 m) and $d_{\rm{AD}} = 2$.

To show the difference between the persistent DIRS-based FPJs with different phase shift distributions,
we consider the following two cases: \emph{Case 1} -- for each DIRS element, the probability of 
choosing phase shift $\pi$/9 is 0.25 
and the probability of choosing phase shift 7$\pi$/6 is 0.75; 
\emph{Case 2} -- each phase shift is equally likely. 
The following benchmarks are compared: the legitimate AP uses the ZF precoder without 
a DIRS-based jamming attack (W/O Jamming), i.e., no DIRS;
the legitimate AP is jammed by the persistent DIRS-based FPJ while the random DIRS phase shifts follow the distributions in \emph{Case 1}  (W/O AJP \& C1) and in \emph{Case 2} (W/O AJP \& C2);
the legitimate AP adopts the AJP for \emph{Case 1} (W/ AJP \& C1) and \emph{Case 2} (W/ AJP \& C2);
the legitimate AP suffers from an AJ with -4 dBm jamming power (AJ w/ $P_{\rm J}$ = -4 dBm), where the AJ is deployed at (-2 m, 0 m, 5 m).

\begin{figure}[!t]
    \centering
    \includegraphics[scale=0.58]{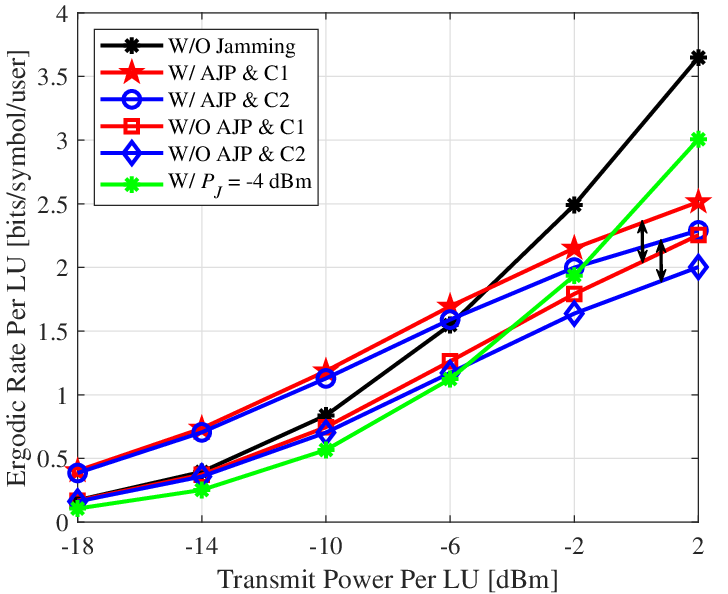}
    \caption{Ergodic rate vs transmit power for different benchmarks under attacks launched by the persistent DIRS-based FPJ.}
    \label{fig4}
\end{figure}

Fig.~\ref{fig4} illustrates the relationship between the rate per LU~\cite{MyGC23,MyGC23Extension} 
under the persistent DIRS-based fully-passive jamming attacks and the transmit power per LU (i.e., $\frac{P_0}{K}$).
From Fig.~\ref{fig4}, we can see that the persistent DIRS-based FPJ can effectively impair the LU rate with neither jamming power nor LU CSI. 
 Specifically, the persistent DIRS-based FPJ 
in \emph{Case 1} and \emph{Case 2} reduces the rate per LU by 28\% and 34.6\% at -2 dBm transmit power, respectively. 
As the transmit power increases, the jamming impact of the persistent DIRS-based FPJ gradually becomes stronger and eventually exceeds that of the AJ. 
 Therefore, we can see that the rates per LU for W/O AJP \& C1 and W/O AJP \& C2 are even worse than that of AJ w/ $P_{\rm J}$ = -4 dBm 
when the transmit power is greater than -6dBm.
The traditional AJ approach requires significant jamming power, 
and increasing the AP transmit power can mitigate the AJ attacks. 
However, increasing the transmit power not only fails to mitigate the jamming impact of the persistent DIRS-based FPJ but even aggravates it.

Compared to the rates per LU obtained from W/O Jamming, the results for 
W/ AJP \& C1 and W/ AJP \& C2 are
better in the low power domain. 
This is because the proposed AJP can to some extent exploit the signals transmitted 
through the DIRS-jammed channels to improve performance. Many practical MU-MISO systems using low-order modulations, such as quadrature phase shift keying (QPSK), can work in the low transmit power domain. 
Moreover, the AJP can mitigate the DIRS-based jamming attacks with different phase shift distributions (e.g., \emph{Cases 1} and \emph{2}).
 Specifically, the AJP in \emph{Case 1} and \emph{Case 2} improves the rate per user
by 19.2\% and 21.1\% at -2 dBm transmit power, respectively. 

\begin{figure}[!t]
    \centering
    \includegraphics[scale=0.58]{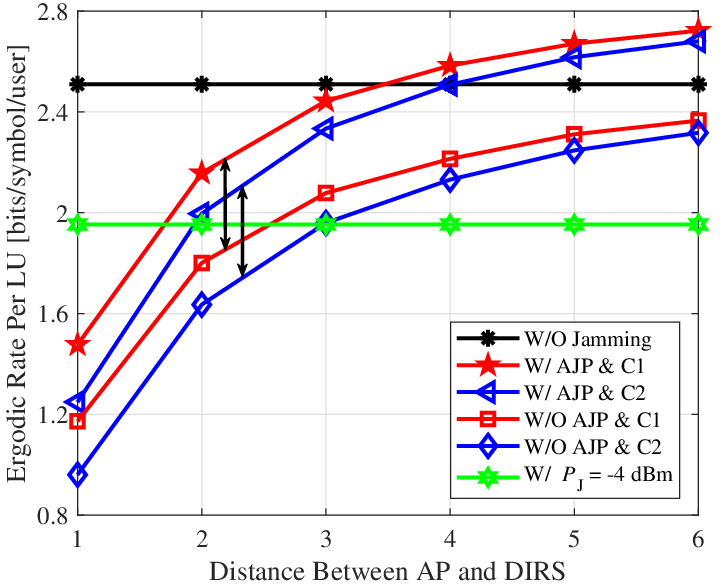}
    \caption{Relationship between the ergodic rate per LU and the AP-DIRS distance for different benchmarks under jamming attacks launched by the persistent DIRS-based FPJ at -2 dBm transmit power per LU.}
    \label{fig6}
\end{figure}

Fig.~\ref{fig6} illustrates the influence of the AP-DRIS distance $d_{\rm{AR}}$. 
As $d_{\rm{AR}}$ increases, the large-scale fading also increases, 
and the jamming impact of the persistent DIRS-based FPJ  is weakened.
The anti-jamming precoder can achieve a rate similar to the case without jamming when the jamming impact is weak.
More Specifically, the results from the AJP, i.e., W/ AJP \& C1 and W/ AJP \& C2, 
achieve better performance when $d_{\rm{AR}} > 4$. 
This is because the gain obtained from the DIRS-based channels using the anti-jamming precoding is greater than the degradation due to the persistent DIRS-based FPJ.
However, the results of W/O AJP \& C1 and W/O AJP \& C2 are always lower than the rates of W/O Jamming.

\section{Future Directions}\label{FutDir}
Based on our investigations, we further outline the following research directions.

\textbf{DIRS-based ACA:} An attacker that wants to achieve a sufficient jamming impact from a
persistent DIRS-based FPJ must ensure that the signals in the DIRS-jammed channels are 
sufficiently strong.
Based on our observations, it is possible that an attacker using a persistent DIRS-based FPJ 
can enhance its jamming impact as follows: 
\begin{enumerate}
\item The attacker can employ a DIRS with one-bit reflective elements whose reflection gain is 
as large as possible.
    Furthermore, the attacker can deploy multiple illegitimate IRSs that use the DISCO approach.
    However, a corollary question is how does the attacker control the phase shifts of all 
    DIRSs in a coordinated manner to maximize the jamming impact? 
\item The attacker can optimize the DIRS phase shift distribution to improve 
    its jamming impact, since we have shown that the performance of a persistent DIRS-based FPJ can vary 
    with different phase shift distributions. 
    However, what is the optimal phase distribution?
\item The attacker can use an active IRS~\cite{ARIS} to replace the passive DIRS 
    to cope with the multiplicative large-scale channel fading. 
    However, optimizing the active IRS gains is challenging because 
    the attacker has no knowledge of the LU channels.
\end{enumerate}

\textbf{Anti-jamming Precoder:} 
Jamming attacks and their anti-jamming strategies can be seen as a form of  ``hand-to-hand combat", 
where each side is constantly trying to gain the upper hand. 
For the proposed AJP, 
our investigation has the following important implication:
as long as the intensity of 
the DIRS-based ACA interference relative to the transmit signals 
can be suppressed below a certain threshold value, 
the persistent DIRS-based FPJ does not degrade the performance of the MU-MISO system, 
but enhances it due to the proposed AJP. 
This suggests that the legitimate AP should minimize the amount of DIRS-based ACA interference
 relative to the strength of the useful signals, and then use the AJP 
 against the persistent DIRS-based FPJ. 

One possible approach to suppressing the strength of DIRS-based 
ACA interference relative to the transmit signals is to introduce legitimate IRSs to 
enhance the desired signals and reduce the relative impact of the DIRS-based ACA interference.
However, DIRS-based ACA interference is also generated  
from the legitimate IRS-related channels. Therefore, a precoding strategy for legitimate IRSs
that significantly enhances the desired signals and does not significantly 
enhance the DIRS-based ACA interference is needed. In addition, methods for detecting the presence of DIRS-based fully-passive jamming attacks 
and their detection probability and false-alarm probability performance should be investigated.

\section{Conclusions}\label{Conclu}
To raise concerns about the potential threats posed by illegitimate IRSs, 
we presented a persistent DIRS-based FPJ that can be implemented using a simple one-bit IRS. 
By introducing significant ACA interference, the persistent DIRS-based FPJ can launch significant 
fully-passive jamming attacks on LUs with neither jamming power nor LU CSI. 
To address the significant threats posed by a persistent DIRS-based FPJ, an AJP 
has been developed that exploits only the statistical characteristics of the DIRS-jammed channels 
instead of their instantaneous CSI. 
A data frame structure that can be used by the legitimate AP to estimate the statistical characteristics 
has also been designed.
The simulation results show that the DIRS-based FPJ with different phase shift distributions (i.e., \emph{Case 1} and \emph{2}) reduces the rate per LU by 28\% and 34.6\% at -2 dBm transmit power, 
but the AJP can improve the rate per user by 19.2\% and 21.1\%, respectively.

\end{document}